\def\cm{\,{\rm cm}}

\def\microK{\,\mu{\rm K}}

\def\km{\,{\rm km}}

\def\kms{\,{\rm km\,s^{-1}}}

\def\Mpc{\,{\rm Mpc}}

\def\s{\,{\rm s}}
\def\d{\,{\rm d}}

\def\aua#1#2{{ #1, }{A\&A,}{ #2}}

\def\apj#1#2{{#1, }{ApJ,} { #2}}

\def\mnras#1#2{{#1, }{MNRAS,} { #2}}

\def\araa#1#2{{#1, }{ARA\&A,} {#2}}

\def\Atoday{\ifcase\month\or
  January\or February\or March\or April\or May\or June\or
  July\or August\or September\or October\or November\or December\fi
  \space\number\day, \number\year}
\def\Etoday{\number\day\space\ifcase\month\or
  January\or February\or March\or April\or May\or June\or
  July\or August\or September\or October\or November\or December\fi
  \space\number\year}

\immediate\write16{... Special symbols are defined}
\def\la{\mathrel{\hbox{\rlap{\hbox{\lower4pt\hbox{$\sim$}}}\hbox{$<$}}}}
\def\ga{\mathrel{\hbox{\rlap{\hbox{\lower4pt\hbox{$\sim$}}}\hbox{$>$}}}}
\def\lse{\mathrel{\hbox{\rlap{\hbox{\raise4pt\hbox{$\<$}}}\hbox{$\simeq$}}}}
\def\gse{\mathrel{\hbox{\rlap{\hbox{\raise4pt\hbox{$\>$}}}\hbox{$\simeq$}}}}
\def\loa{\mathrel{\hbox{\rlap{\hbox{\lower4pt\hbox{$\approx$}}}\hbox{$<$}}}}
\def\goa{\mathrel{\hbox{\rlap{\hbox{\lower4pt\hbox{$\approx$}}}\hbox{$>$}}}}

\def\ed{\end{document}}

\def\beq#1{\begin{equation}\label{#1}}
\def\eeq{\end{equation}}
\def\beqa#1{\begin{eqnarray}\label{#1}}
\def\eeqa{\end{eqnarray}}

\def\bfig{\begin{figure}[h] \centerline{\hbox{}}\vfill}
\def\efig{\end{figure}\vfill\newpage}

\def\spose#1{\hbox to 0pt{#1\hss}}
\def\simlt{\mathrel{\spose{\lower 3pt\hbox{$\mathchar"218$}}
     \raise 2.0pt\hbox{$\mathchar"13C$}}}
\def\simgt{\mathrel{\spose{\lower 3pt\hbox{$\mathchar"218$}}
     \raise 2.0pt\hbox{$\mathchar"13E$}}}
\def\simpropto{\mathrel{\spose{\lower 3pt\hbox{$\mathchar"218$}}
     \raise 2.0pt\hbox{$\propto$}}}

\documentstyle[12pt,cmb,psfig]{article}
\begin{document}
\heading{COBRAS/SAMBA AND MEASUREMENTS OF THE SUNYAEV-ZELDOVICH EFFECT}

\author{Martin G. Haehnelt$^{1}$} {$^{1}$ Max-Planck-Institut 
        f\"ur Astrophysik, Karl-Schwarzschild-Stra\ss e 1, 
        85740 Garching, Germany.}

\begin{abstract}{\baselineskip 0.4cm 
The recently approved COBRAS/SAMBA cosmic microwave background (CMB)
mission  will also have a major impact on measurements of the 
Sunyaev-Zeldovich (SZ) effect. The frequency of three of the channels 
(150 GHz, 217 GHz, 353 GHz)  are chosen to optimize measurements of 
the thermal and kinetic SZ effect mainly caused by  hot gas in 
clusters of galaxies.  Estimates  of the number of detected clusters  
are somewhat uncertain due to incomplete knowledge of the gas mass
function of  clusters and the distribution of hot gas at large radii. 
Straightforward interpolation of X-ray observations of the 200  
X-ray brightest clusters gives a firm lower limit of  $\sim 3000$ 
detected clusters of which roughly half should be resolved. If the gas 
distribution of clusters should turn out to be favourable and/or the 
cluster evolution with redshift is weak, this number could be higher 
by a factor of up to ten. Using an optimal filtering technique a
measurement of the bulk peculiar velocity of a sample of about 
200 of the resolved clusters will be possible with a 
$1\sigma$ error of $100-200 \km\s^{-1}$.
}
\end{abstract}

\section{Sensitivity of COBRAS/SAMBA}

\begin{figure}[t]
\centerline{
\psfig{file= 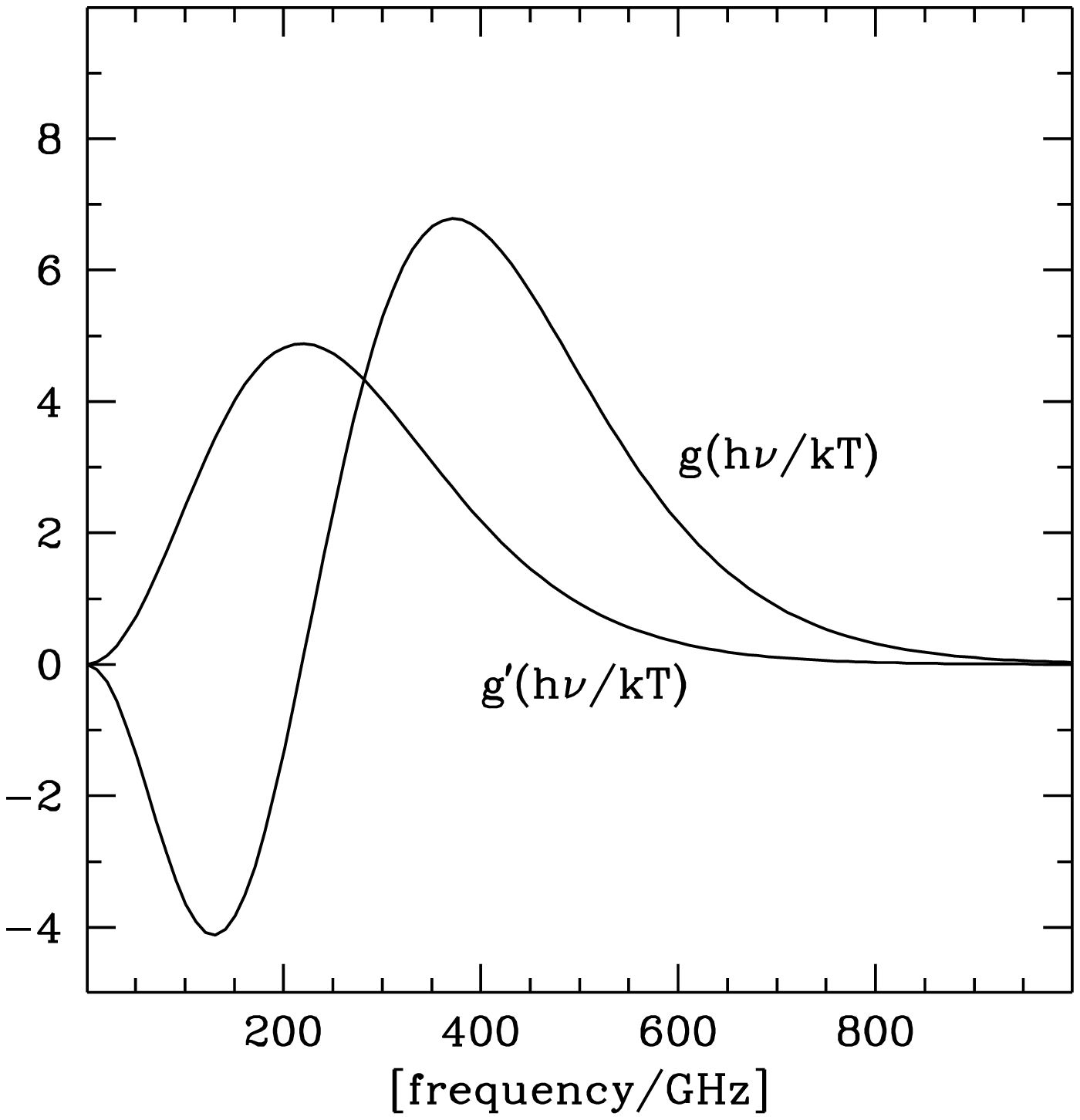,width=8.5cm,angle=0.}
\psfig{file= 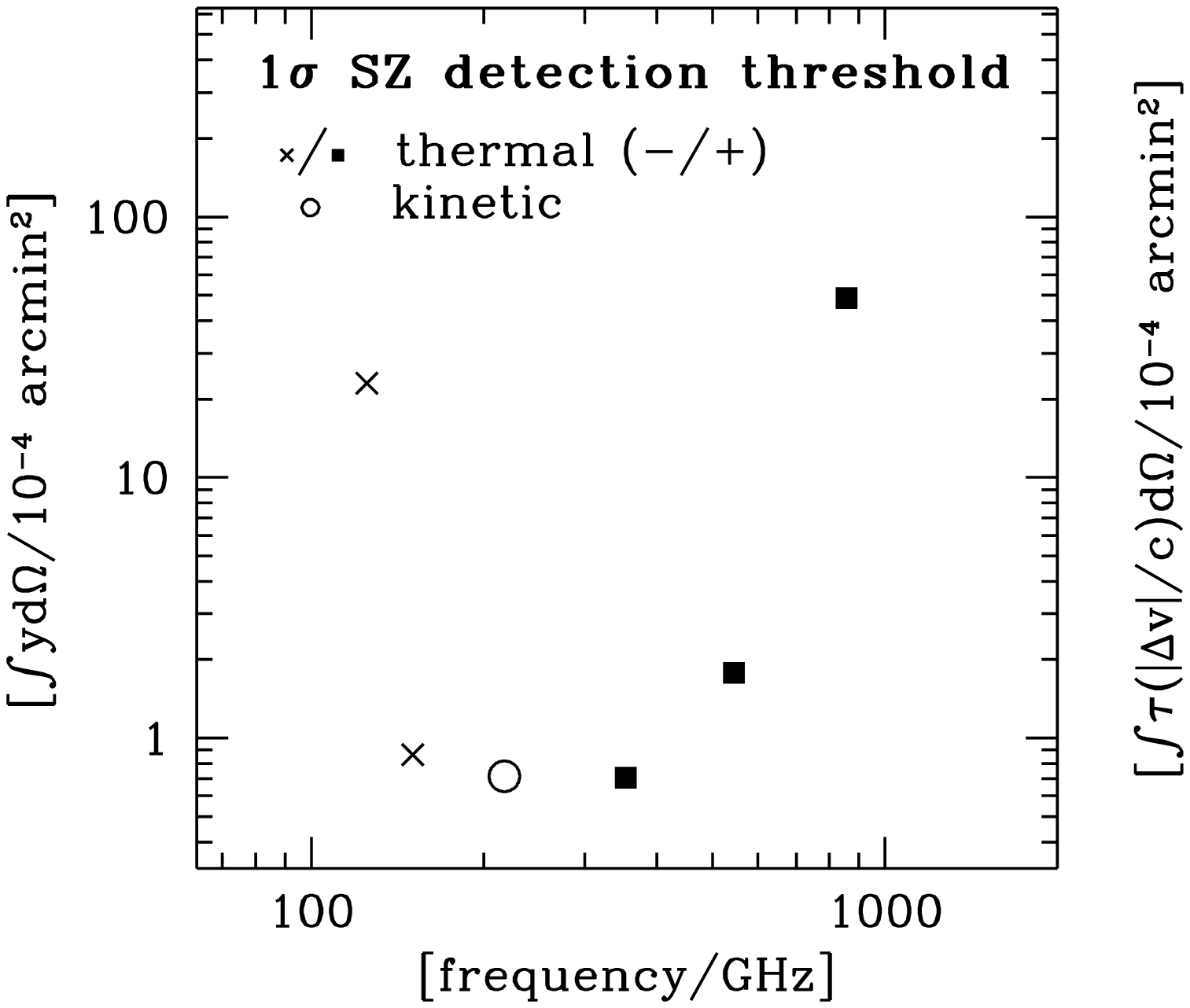,width=9.5cm,angle=0.}
}
\vspace{-1.5cm}
\caption{\baselineskip 0.4cm 
The left panel shows the frequency dependence of the thermal ($g$) and 
kinetic SZ ($g^{\prime}$) effect (equation 1 and 2). In the  right panel  
the  $1\sigma$ SZ detection threshold of the high 
frequency channels is shown in terms of the integrated Compton 
parameter $Y= \int{ y \d \Omega}$.}
\end{figure}

For an unresolved cluster  the ``SZ-flux'' due to Compton scattering 
by hot electrons can be written as 
\beq{DeltaIeq}
         S_{\nu} =  f_{\nu}\; g(h\nu/kT) \;  Y , 
\eeq
where $g(h\nu/kT)$ describes  the frequency dependence of the  Compton 
distortion (Fig. 1), $f_{\nu}$ is a normalization constant  
and $Y= \int{ y \d \Omega}$ is the Compton y parameter integrated 
over solid angle \cite{Suna} \cite{Repha} \cite{Rephb}. 
$Y$  is proportional to 
the typical mass times the   mass-weighted temperature within the 
radius where the density profile of the cluster becomes steeper than 
isothermal (and/or the gas temperature starts to drop). 
The crosses and squares in Figure 1b  show the   $1\sigma$ SZ detection 
threshold of the high frequency channels
in terms of $Y$ . 
A typical value for the ``SZ-channels'' at 150 and 353 GHZ 
(maximum negative/positive SZ-flux)  is $10^{-4} {\rm arcmin}^2$. 
This takes only detector noise into account and will be somewhat 
increased by residual noise from the foreground subtraction
procedure. $3-10 \times 10^{-4} {\rm arcmin}^2$ should be considered 
as a reasonable range for the expected detection  threshold of 
an unresolved cluster (marked by the shaded region in Figure 2.)

\section{SZ number counts of unresolved clusters}

\begin{figure}[t]
\centerline{
\psfig{file= 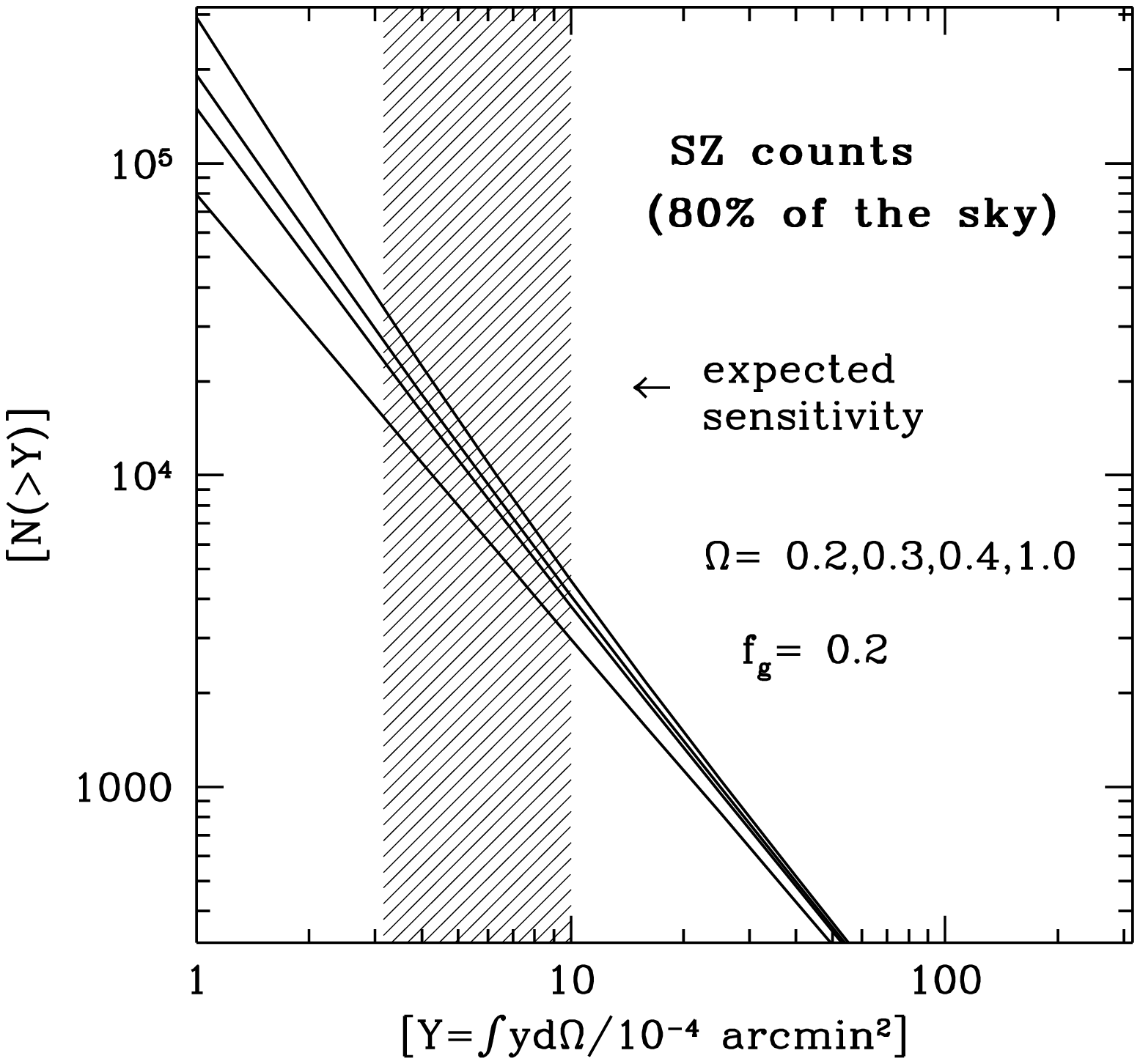,width=8.5cm,angle=0.}
\psfig{file= 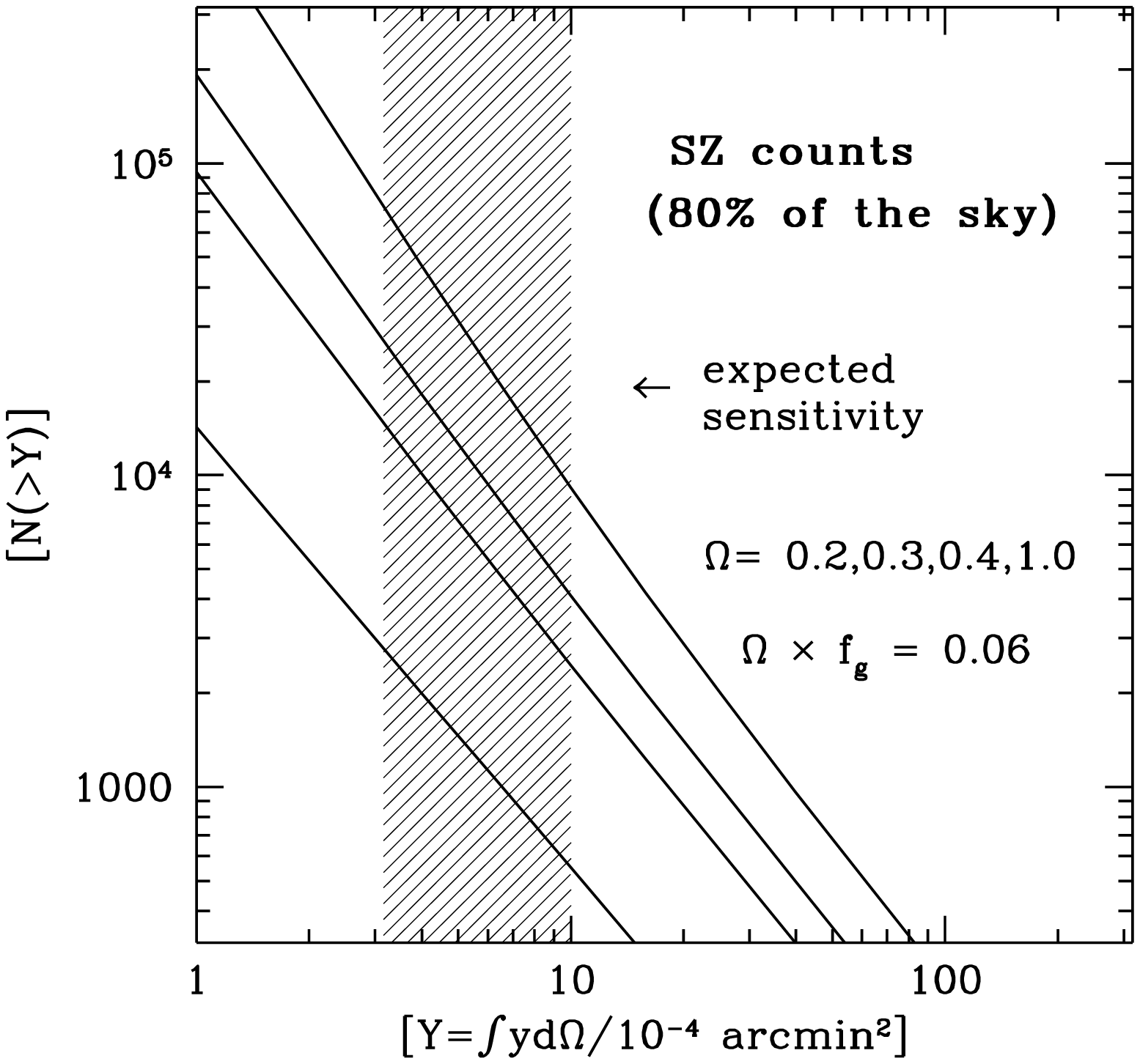,width=8.5cm,angle=0.}
}
\vspace{-1.5cm}
\caption{\baselineskip 0.4cm  
         The expected number of detected clusters is plotted as 
         function of the integrated Compton parameter $Y= \int{ y \d
         \Omega}$ for different values of the total matter 
         density $\Omega$ as indicated ($\Omega$ decreasing upwards).  
         The present-day mass function of clusters was evolved
         backwards  in time using the Press-Schechter 
         formalism. An spectral index of $n=-1$ for the primordial 
         density fluctuation spectrum on
         cluster scales is assumed.
         The expected sensitivity of COBRAS/SAMBA is indicated by the 
         shaded area. The left panel is for fixed gas mass fraction 
         $f_{\rm g} = 0.2$ , while the right panel assumes compatibility 
         with the nucleosynthesis constraint 
         $\Omega \times f_{\rm g} = 0.06$.}

\end{figure}

Straightforward interpolation of the properties of a flux-limited 
sample of the 200 X-ray  brightest clusters \cite{Ebe} by a factor 
3-5 in flux 
gives a firm lower limit of  $\sim 3000$ detected clusters. More
accurate estimates are difficult because the typical gas mass 
and mass-weighted temperature within the radius where the density
profile of the cluster becomes steeper than isothermal 
(and/or the gas temperature starts to drop) are 
difficult to determine from  X-ray observations which 
generally probe the gas and temperature distribution at 
considerably  smaller radii. The typical  Y-parameter of clusters of 
a given present-day  number density  is probably uncertain by a 
factor of two. The estimated number counts are also sensitive to
\begin{itemize}
\item 
the exact shape of the present-day mass function, especially 
the slope at  large masses,
\item
the details of the assumed evolution of the cluster mass/temperature
function,   
\item
and the cosmological parameter. 
\end{itemize}
Nevertheless, the Press-Schechter formalism can be used to evolve the  
observed present-day mass function of clusters  backward 
in time to get a feeling for the influence of the uncertain 
parameter \cite{Cen} \cite{deLuca}. Assuming the usual 
observationally normalized
scaling of the cluster temperature $T\propto M^{2/3}\,(1+z)$
we can work out the expected  number counts $N(>Y)$.  
In Figure 2 these are shown for different values of $\Omega$ and 
different assumptions for the  total fraction of hot gas.  The left panel 
assumes a fixed gas mass fraction while the right panel assumes a 
gas mass fraction compatible with the nucleosynthesis constraint.
The total number of clusters detected 
by COBRAS/SAMBA will depend strongly on the assumed sensitivity 
($N\sim Y^{-3/2}$). The large difference between the two panels 
illustrates the  uncertainties due to our insufficient knowledge 
of the total mass of hot gas in clusters.

\section{Properties of the detected cluster sample} 

\begin{figure}[t]
\centerline{
\psfig{file= 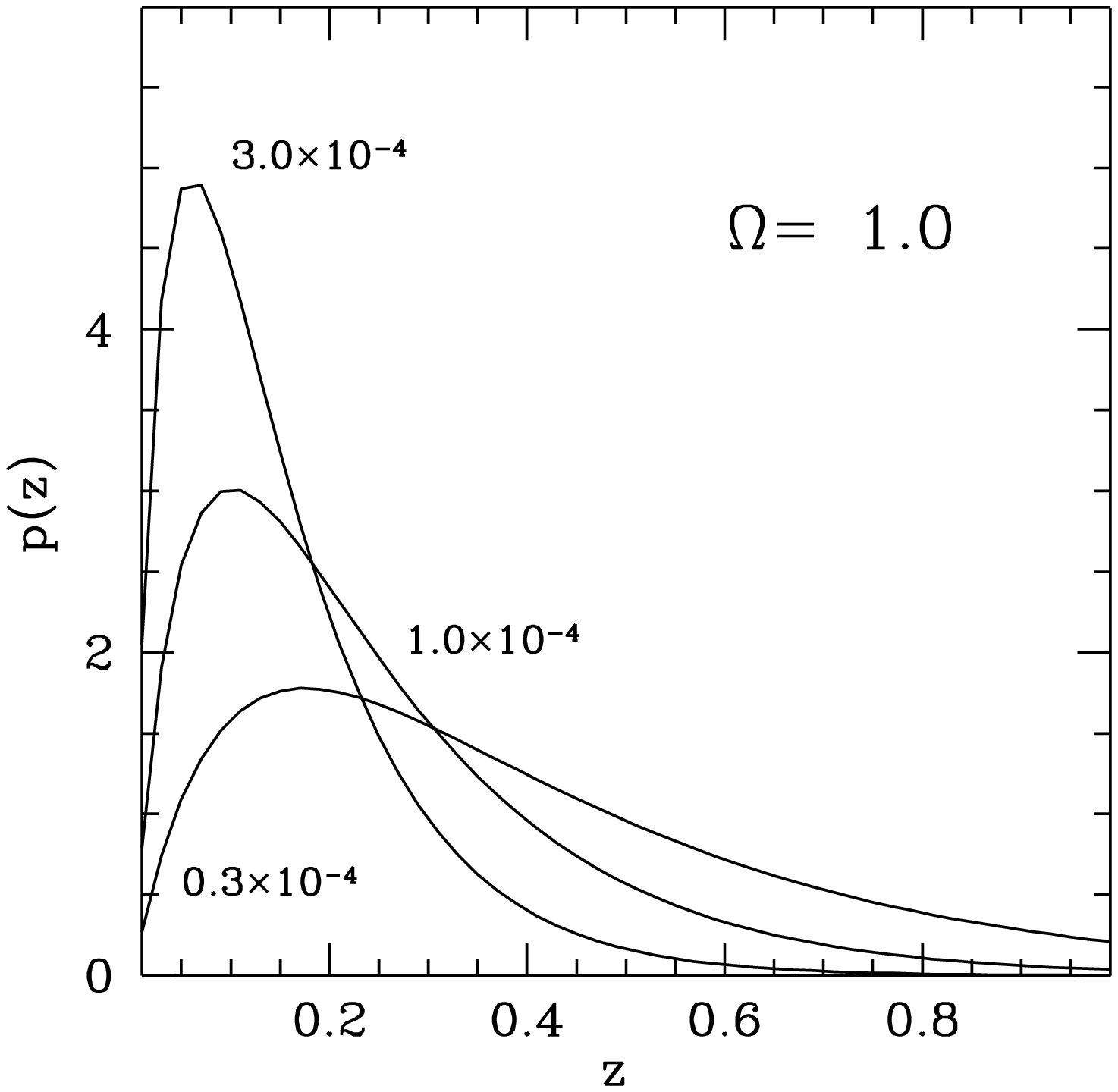,width=8.5cm,angle=0.}
\psfig{file= 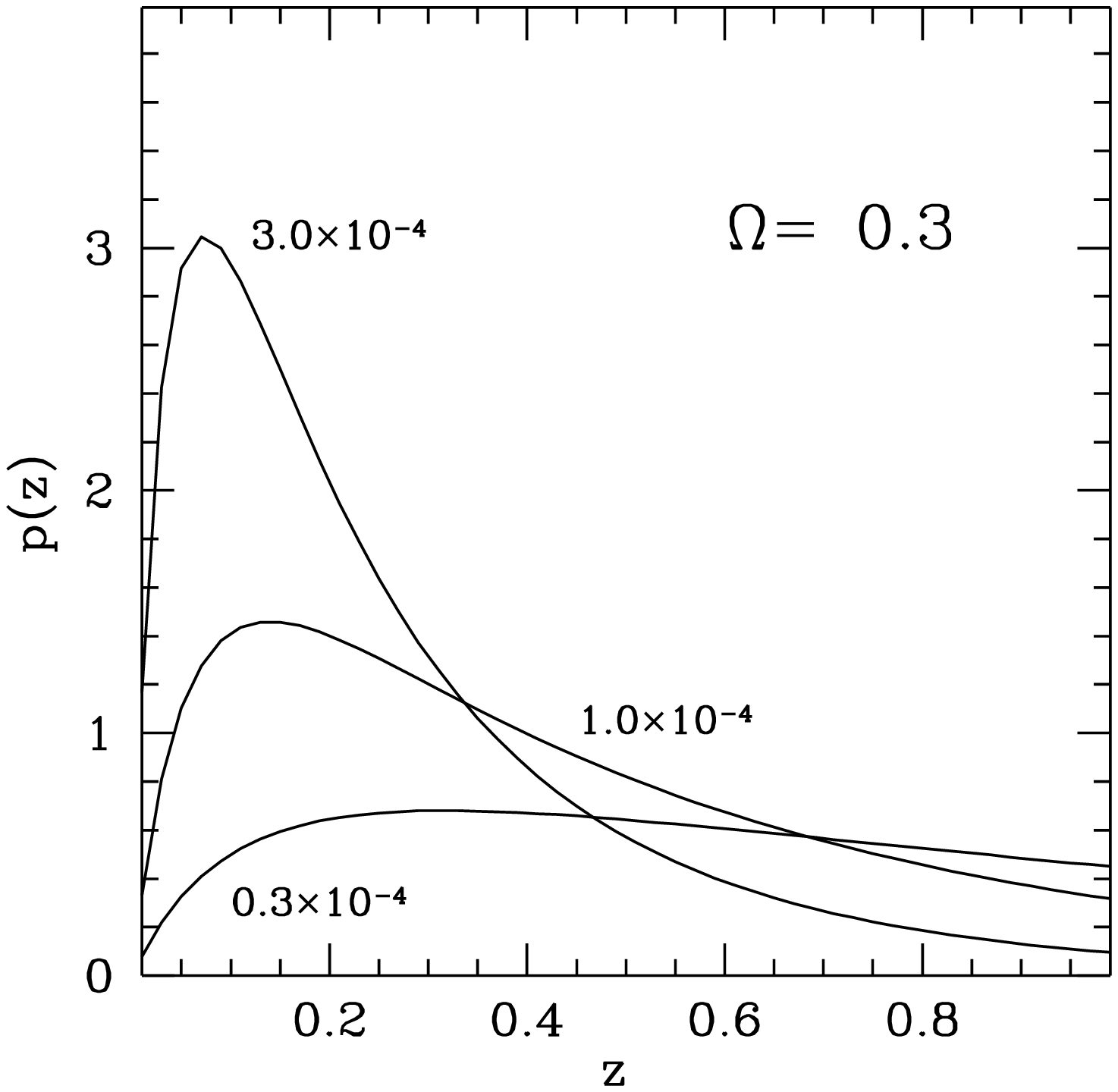,width=8.5cm,angle=0.}
}
\vspace{-1.0cm}
\caption{\baselineskip 0.4cm 
The left panel shows the expected redshift distribution for $\Omega=1$
and for different sensitivity limits in terms of the integrated 
Compton parameter $Y= \int{ y \d \Omega}$ as indicated on the plot. 
The right panel is the same for $\Omega=0.3$.}
\end{figure}

The properties of the detected cluster sample 
will also depend on the uncertainties mentioned above.
If  a detection threshold of  $3 \times 10^{-4} {\rm
arcmin}^2$ and $\Omega =1$ is assumed a sample 
of $10^{4}$ clusters is expected. This is   
about a factor of two larger than 
the number of clusters  in the  extended Abell 
catalogue. As shown in Figure 3a the typical redshift 
would be $z=0.1$,  very similar to that in the Abell 
catalogue, while the typical mass would be somewhat smaller. 
Unfortunately, the sensitivity and spatial resolution of 
COBRAS/SAMBA will not be sufficient to make use of the fact 
that the ``surface brightness'' of the SZ effect is independent 
of redshift. Figure 3b shows that a flat redshift 
distribution would only be seen with considerably better 
sensitivity and in a low $\Omega$ universe. However, a high redshift
tail might still be seen if $\Omega$ is low \cite {Bar}.  
 
\vfill 
\break

\section{The kinetic SZ effect} 

\begin{figure}[t]
\centerline{
\psfig{file= 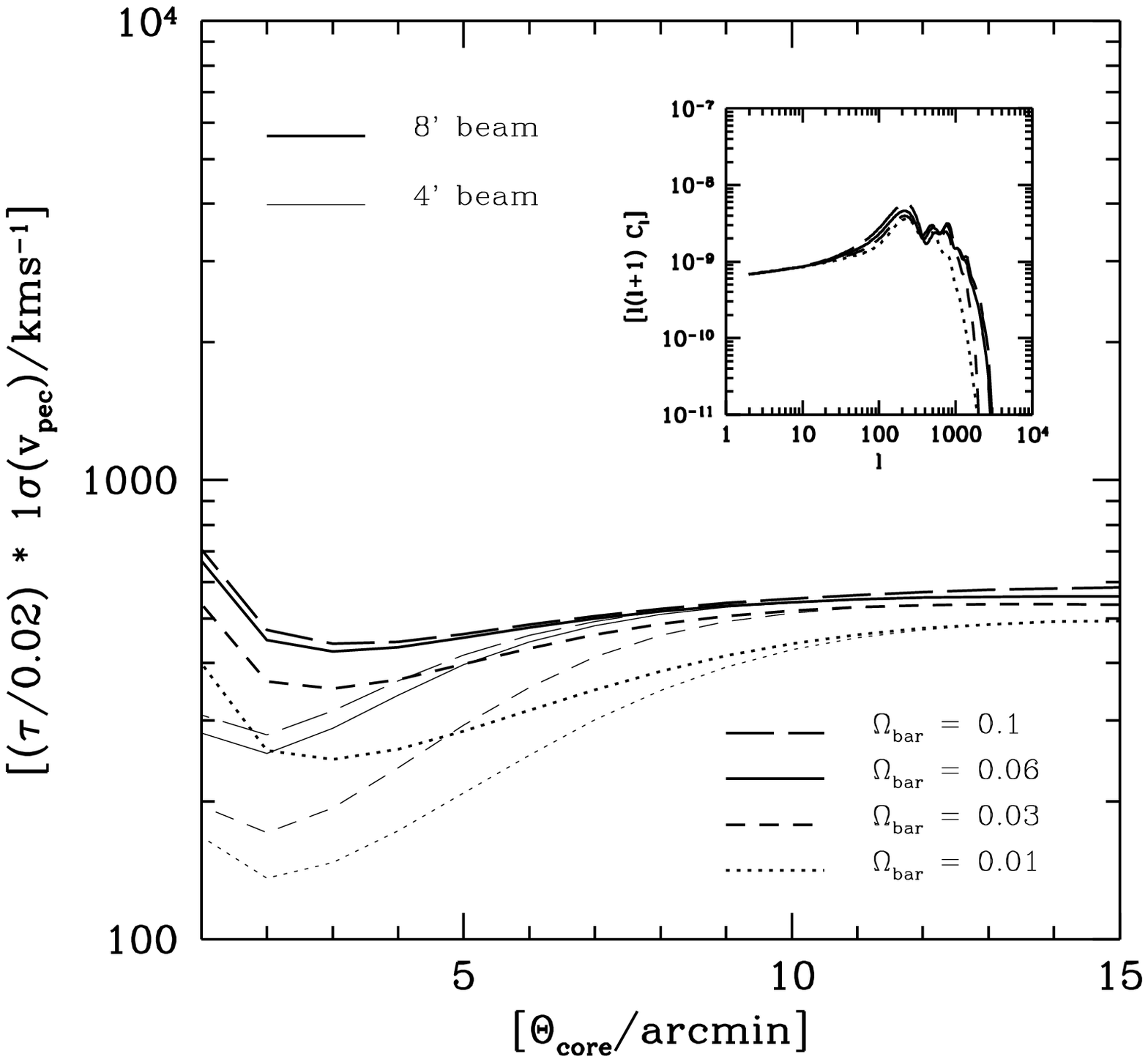,width=8.5cm,angle=0.}
\psfig{file= 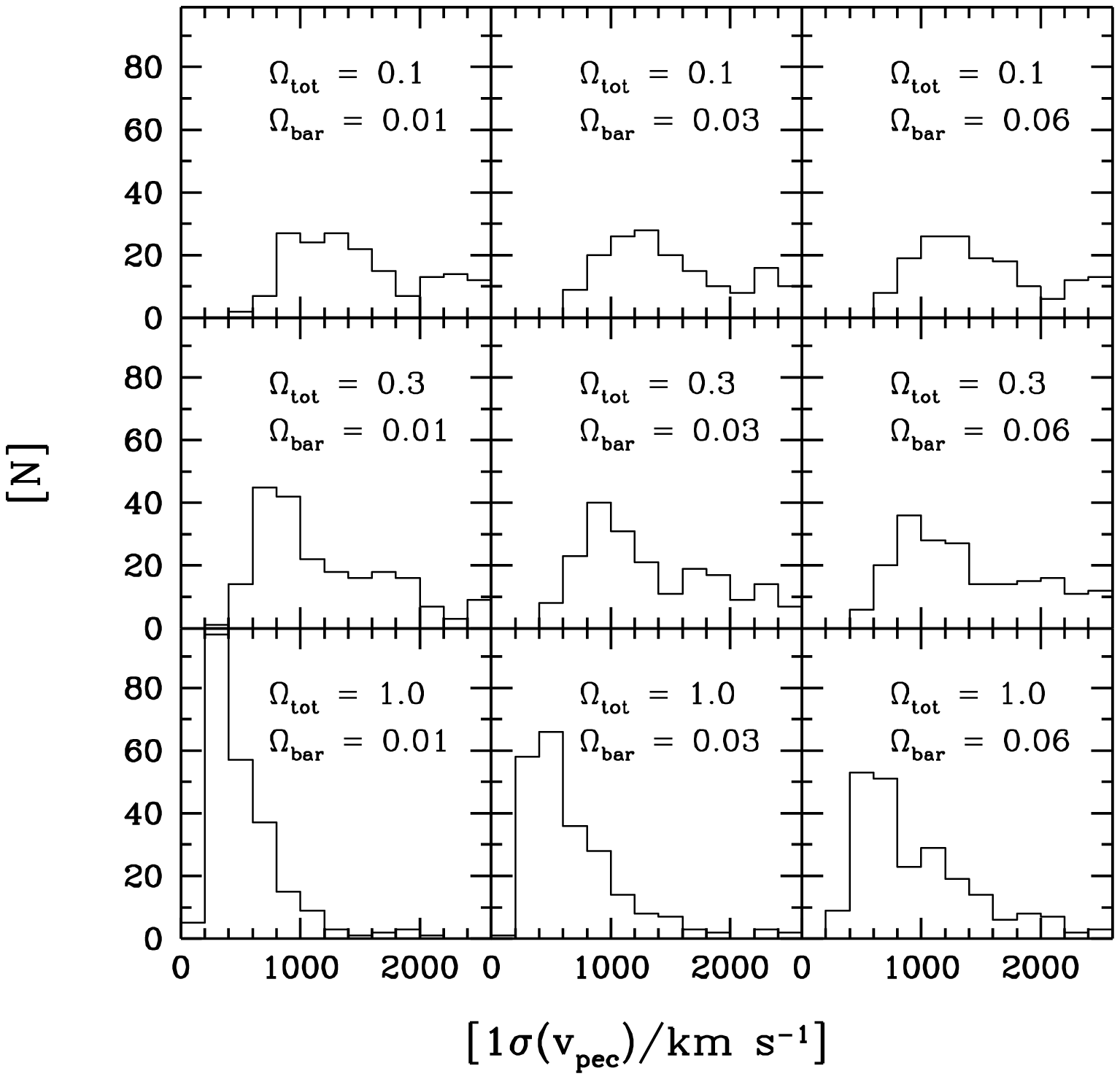,width=8.5cm,angle=0.}
}
\vspace{-1.5cm}
\caption{\baselineskip 0.4cm 
         The  left panel shows the $1 \sigma$  error in the determination 
         of the peculiar velocity as a function of the core radius of 
         the cluster using an axisymmetric ``optimal''  filter 
         function for a standard CDM scenario ($\Omega_{\rm tot} =1$, 
         $ H_{0}= 50 \km \s^{-1} \Mpc^{-1}$) with varying baryonic 
         fraction. The pixel noise is fixed and corresponds to
         $7\microK$ in the  4' (FWHM) beam. Thick curves are for a beam size 
         of 8' and thin curves are for a beam size of 4'.
         The inset shows the angular power spectra of temperature 
         fluctuations for the three cosmological models. 
         The right panel shows the distribution of expected $1\sigma$ errors 
         of peculiar velocity measurements for the XBACs cluster sample  
         for different cosmological models . $\Omega_{\rm tot}$ 
	 and $\Omega_{\rm bar}$ vary as indicated in the plot. An axisymmetric 
         optimal filter was applied using a $\beta$-model for 
         the cluster gas distribution.}
\end{figure}

For an unresolved cluster  the kinetic ``SZ-flux'' due to peculiar
motion of the hot gas in a cluster with respect to the CMB
can be written as 

\beq{DeltaIeqa}
         S_{\nu} =  f_{\nu}^{\prime}\; g^{\prime}(h\nu/kT) \; 
         \int{\tau |v_{\rm pec}/c|\d \Omega}, 
\eeq
where $g^{\prime}(h\nu/kT)$ describes  the frequency 
dependence of the  kinetic  distortion (Fig. 1a), 
$f_{\nu}^{\prime}$ is a normalization constant,  
$\tau$ is the Thompson optical depth and $v_{\rm pec}$ is  the peculiar 
velocity of the cluster. The frequency dependence is that of a
temperature fluctuation in the CMB and for  a typical cluster 
the change in brightness temperature at the 
cluster center is of order 
\beq{DeltaTeq}
        {\Delta T} \sim    
             30 \;
       \biggl ( {n_{\rm e} \over 3\times 10^{-3} \cm^{-3} } \biggl )\;
       \biggl ( {r_{\rm c} \over 0.4 \Mpc} \biggl )\;
       \biggl ( {v_{\rm pec} \over 500 \km \s^{-1}} \biggl )\;
          \microK,
\eeq
where $n_{e}$ is the electron density in the core, $r_{c}$ is the
core radius and we have scaled to the values of the Coma cluster 
(assuming a distance of 140 Mpc) \cite{Sunb} \cite{Repha}. 
The open circle in Figure 1b shows the
sensitivity of COBRAS/SAMBA  for an unresolved cluster in terms of 
$\int{\tau |v/c|\d \Omega}$.

\section{Measuring bulk velocities of clusters}

For the kinetic effect the expected noise level due to confusion
with primary fluctuations is generally of order or larger than the
expected  signal.  It is therefore essential to use the knowledge 
of the CMB ``noise" properties and the gas distributions of the 
individual clusters (which can be obtained by the mission 
itself and from X-ray observations, respectively).  
This knowledge makes it possible to analyze the CMB maps 
of  resolved clusters with a spatial
filter optimized for individual clusters. An improvement in
signal-to-noise  by a factor of two is easily achievable, and even a 
factor of 10 is  possible if the gas mass  distribution is well known 
from a high-quality X-ray map \cite{Hae}. 
The final signal-to-noise ratio depends crucially on the cosmological
model and the angular resolution of 
the instrument (Fig. 4a), and it is not currently clear 
whether a meaningful peculiar velocity measurement for individual 
clusters will be possible.
Prime candidates are X-ray luminous clusters 
at intermediate redshift 
with core radii just below the Doppler peak scale. 
For a favourable but still rather standard cosmological 
scenario (standard CDM with low baryon fraction) and a good 
angular resolution  (4' FWHM), the peculiar velocity of as
many as 30 individual clusters might be determined accurately. 
Even if this is impractical, it should still be possible 
to  determine the bulk motion of an ensemble of 200 X-ray 
luminous clusters at redshifts $\ga 10 000 \km \s^{-1}$ 
with an accuracy of order $100-200 \kms$ (Fig. 4b).

\section{Concluding remarks}

COBRAS/SAMBA will produce  the first all-sky SZ maps. 
In these maps most known clusters of galaxies  will be 
detected roughly half of which will be resolved. It will therefore 
be possible  to study the gas distribution of a large sample of
clusters  out to the virial radius. Once the  cosmological parameter 
are known with good precision (the primary aim of the COBRAS/SAMBA 
mission)  problems like a possible clumping of the gas can be addressed. 
Relativistic corrections to equation (1) and (2) might also allow to 
achieve accurate  measurements of the  temperature of the cluster 
gas \cite{Rephb} \cite{Gia}. 
Comparison of  SZ and X-ray properties of clusters 
will significantly advance our understanding of cluster evolution. 
If the gas mass fraction in clusters is high and there is as little 
evolution in the cluster population as indicated by current X-ray 
observation the SZ effect might also be a suitable tool 
to detect clusters at high redshift in appreciable numbers. 
So far, not much is known 
about bulk velocities on  scales larger than $5 000$ to 
$10 000 \km\s^{-1}$. The measurement of the bulk  velocity of a 
volume-limited sample of  119 Abell clusters  out to a distance of 
$15 000\kms$ gave a value considerably higher than 
expected in most cosmological scenarios \cite{Pos}.
The completely independent measurement by COBRAS/SAMBA  
on even  larger scales should clarify this situation.

\vfill
\end{document}